# A Perspective on Future Research Directions in Information Theory

Members of "Future Directions" Committee
Jeff Andrews, UT Austin
Alex Dimakis, UT Austin
Lara Dolecek, UCLA
Michelle Effros, Caltech
Muriel Medard, MIT
Olgica Milenkovic, UIUC
Andrea Montanari, Stanford
Sriram Vishwanath, UT Austin
Edmund Yeh, Northeastern

External Contributors (Section):
Randall Berry, Northwestern University (Finance)
Ken Duffy, National University of Ireland, Maynooth (Genomics)
Soheil Feizi, MIT (Genomics)
Saul Kato, Institute of Molecular Pathology, Austria (Neuroscience)
Manolis Kellis, MIT (Genomics)
Stuart Licht, Biochemistry and Bioanalytics Group, Sanofi-Aventis (Genomics)
Jon Sorenson, Clinical Bioinformatics, InVitae (Genomics)
Lav Varshney, UIUC (Finance)
Haris Vikalo, UT Austin (Genomics)

**Abstract –** Information theory is rapidly approaching its 70th birthday. What are promising future directions for research in information theory? Where will information theory be having the most impact in 10–20 years? What new and emerging areas are ripe for the most impact, of the sort that information theory has had on the telecommunications industry over the last 60 years? How should the IEEE Information Theory Society promote high-risk new research directions and broaden the reach of information theory, while continuing to be true to its ideals and insisting on the intellectual rigor that makes its breakthroughs so powerful? These are some of the questions that an *ad hoc* committee (composed of the present authors) explored over the past two years. We have discussed and debated these questions, and solicited detailed inputs from experts in fields including genomics, biology, economics, and neuroscience. This report is the result of these discussions.

# Table of Contents

# 1   What is Information Theory?

**Definition.** Information theory is a mathematical science that studies the ultimate limits of, and optimal methods and algorithms for:
1. The representation of information;
2. The communication of information;
3. The processing and utilization of information.

Having formally been launched by Claude Shannon of Bell Labs in 1948 in one of the great intellectual accomplishments of the $20^{th}$ century [1], information theory has matured into a field that defies easy categorization. Traditionally residing in departments of electrical engineering due to its origins and historical focus on problems in telecommunications, today it has matured far beyond being a subfield of electrical engineering. Sometimes considered a field of applied mathematics or probability, this classification also fails to capture its richness, for example the vast and unique theory of coding, the characterization and analysis of networks, and its increasing application to many fundamental problems in natural and life science.

The essence of information theory can be thought of as:
- The development of mathematical abstractions and models that capture the salient features of a problem
- A rigorous mathematical analysis that establishes fundamental limits and tradeoffs to the problem, that are agnostic to any particular approach to solving it and thus provide an upper bound or "speed of light" that can be approached but not exceeded
- The drawing of insights and intuition from the analysis and limits, that often leads, eventually, to the development of near-optimal architectures and algorithms

Since "information" plays a central and increasing role in many fields of science and engineering particularly as the generation and storage of information increases exponentially, the information theoretic approach is attractive for many fields beyond communications. This report takes the view – Shannon's warning not-withstanding [2] – that information theory's future lies largely in the deepening and broadening of such boundaries, and also the exploration of new boundaries with additional fields. Information theory's abstract and rigorous framework offers a great deal to many fields where the approach is typically more observational or heuristic. Indeed, soon perhaps, information theory may be seen as an indispensable tool in traditional fields, not unlike calculus or computer programming.

While by no means exhaustive (and making no attempt to cite the vast literature), this report represents an effort to articulate some of these exciting connections – past, current, and future – and begin a discussion on how to maximize the impact of information theory in the coming decades. We begin with native fields for information theory before moving into several more emerging areas. To cope with some of the challenging problems that this report outlines, information theorists will likely need to continue to develop and embrace new abstractions, models, tools, methods, and their own imaginations.

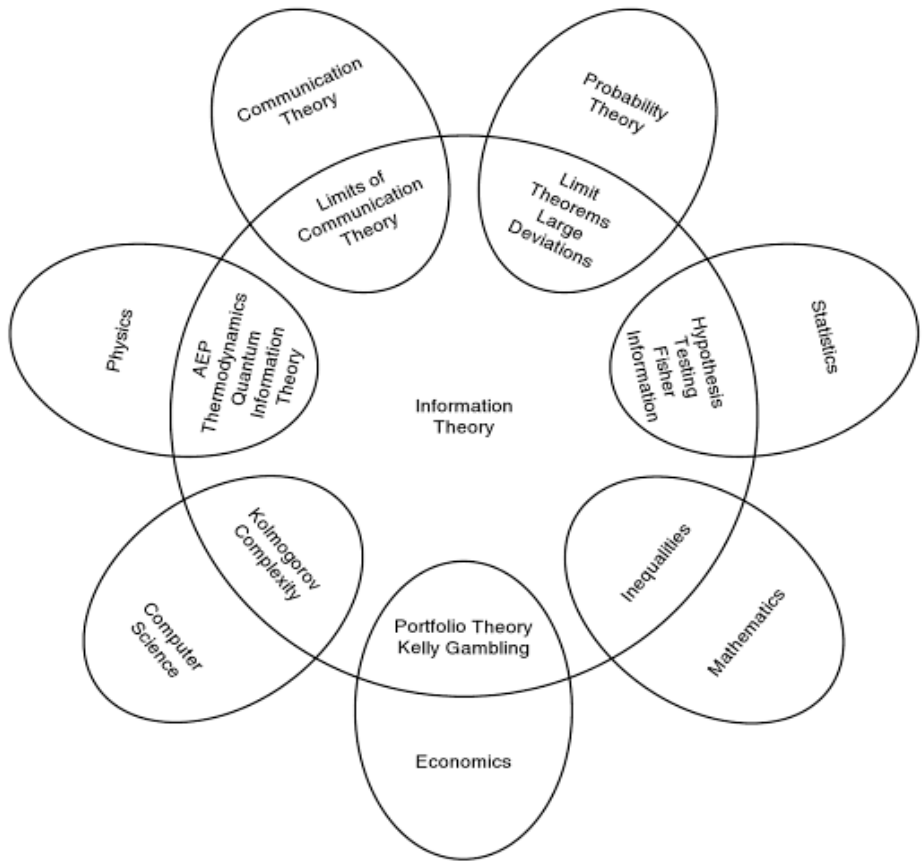

Figure 1: Cover and Thomas's 1991 Representation of the Intersection of Information Theory and Other Fields. This report can be viewed as a more detailed update of such a figure, with descriptions of the intersections [3].

## 2. Communications

Communication, i.e. the transportation of information over space or time (i.e. storage) in the presence of noise and other impairments, is the most widely appreciated and well-developed area of information theory, and the main subject of Shannon's 1948 paper [Sha48]. Important successes include:

- The capacity of noisy, fading, and broadcast/MAC channels
- Development of near-optimal (optimal meaning capacity achieving) codes and decoding algorithms
- A nearly complete understanding of the capacity of interference channels, including achievability schemes such as interference alignment and lattice-based codes
- The application of these and other discoveries to modern consumer communication systems such as cellular and wireless LANs, which to a large extent are based on a blueprint of what information theory prescribes in the case of wideband fading channels.

Because of the great deal of work done over the last several decades, uncharted future directions are challenging to identify for canonical communication models. However, there are several interesting future directions in information theory applied to communication systems. We also refer the reader to the next section on networks, where many unsolved challenges remain.

Connections and Future Directions:
- **Massive MIMO (multi-input multi-output) Communication.** It is becoming increasingly viable to put previously unimagined numbers of antennas on transmitters and possibly receivers, particularly if millimeter wave frequencies (30-300 GHz) begin to be used for terrestrial purposes. Although much classical MIMO information theory applies in this regime, there are many particular cases where that theory is not applicable. For example, channel state information is nearly impossible to acquire for each transmit-receive antenna pair.
- **Relaxations of traditional models:**
  - **Finite Block Length Coding and Capacity.** Shannon's formula relies on several fairly strong assumptions such as infinite block length and vanishing error probability, that nevertheless nearly hold in many practical scenarios. One important relaxation is for short block lengths, due to several short packet scenarios of contemporary interest, e.g. machine-to-machine or super low-latency applications. A major simplification over error exponents was recently discovered in this short packet and finite error rate regime via the channel dispersion and a Gaussian approximation.
  - **Delay-tolerant feedback-based transmission techniques**. One major challenge with many information theory inspired communication techniques, such as dirty paper coding and interference alignment, is that they require accurate and timely channel state information at the transmitter. The delay issue is particularly troubling since it does not scale with bandwidth and in fact becomes larger as a percentage of the round trip time as technology advances. One recent example of novel work to escape the latency deathtrap is the Maddah-Ali and Tse scheme, which uses a novel information theory inspired technique to use arbitrarily delayed feedback while achieving performance that is within a bounded gap from the optimal capacity scaling.
- **Informing and iterating with the telecommunications industry.** Wireless communications is a massive global industry with billions of consumers that demand constant technology innovation. Thousands of PhDs, many in information theory, toil at highly innovative companies to constantly improve the performance of these networks. Information theory has continually abetted this industry including currently by developing reliable and easily digestible new theories that explain complex new phenomenon and allow continued innovation. Although not basic science like some of the topics below, there is no reason to suspect that this crucial role of information theory will diminish in the coming decades. It does demand that information theorists working on communication problems strive to stay informed on the state of the art in industry by, for example, staying up to date on key features of future standards, or regularly conversing with top industry innovators.

# 3 Networks and Networked Systems

Many disciplines are interested in the flow of information in large interconnected systems. First we specifically discuss electronic communication networks, both wired (e.g. the Internet) and wireless (e.g. ad hoc networks), although many of the principles extend to other types of networks. A heterogeneous communication network can be viewed as a collection of simpler homogeneous networks (such as broadcast, multiple access, interference and relay channels) and point-to-point links. Such decompositions are well motivated in wired networks due to physical separation of links, although benefits can be gained by analyzing the collection of links as a whole, most notably via network coding.

Connections and Future Directions:
- **Multi-terminal information theory** has been the core of this area, pushing our fundamental understanding, starting from small building-block networks.
- **Network coding** has been an algorithmic approach to obtaining achievable schemes for different types of network communication problems. More recently, new ways of managing and understanding interference have opened up a new landscape on network communication schemes. Interference does not necessarily happen over the wireless physical medium but can also occur in wired network coding problems (e.g. in the distributed storage systems used in modern data centers).

Beyond communication networks, information theory will find application in other types of large networks. Examples include:
- **Fundamental properties of large graphs**, for example graph-matching in the presence of noise (missing or erroneous edges), de-anonymizing graphs with side information, and fundamental scaling laws of large graphs. This includes a massive variety of applications and problems that can be modeled as graphs.
- **Nano-circuit design**. As integrated circuits become smaller and smaller, their processing will become less and less reliable, due to limits in manufacturing processes (i.e. lithography) and soon, quantum effects. Thus, circuit design will increasingly resemble a coding problem (for channel models that have not been studied in depth in existing coding literature, e.g. processing noise).
- **Reliable data storage.** New storage devices possess large operational spatio-temporal asymmetry, so implementing codes developed for symmetric channels (as in the past) is suboptimal and needlessly adds to the cost of a storage system due to overprovisioning. Asymmetric channels are an important area of research in view of novel storage systems.

# 4 Control Theory

The relationship between information theory and control is a long one, both from the standpoint of control theory as an enabling framework for studying information theoretic problems as well as from the standpoint of information theory as a tool to attack complex control problems.

Connections and Future Directions:
- **Capacity of feedback channels**: The capacity of feedback channels is a classical information theoretic problem, studied by Shannon in the context of discrete memoryless

- channels (DMCs). Although the capacity of feedback single user DMCs is found to be unaffected by feedback, feedback is now known to be useful from a computational complexity, error-rate decay and therefore, overall latency perspective for single user DMCs, as well as in enhancing capacity in multiuser and channel-with-memory cases. In a vast majority of existing literature on this topic, control theory has played an essential role in developing new achievability strategies for feedback channels.
- **Cognitive radios and Dynamic Spectrum Access:** Communication and control come together in understanding mechanisms for cognition in communication systems, particularly for dynamic spectrum access, where "state" plays a major role, since nodes attempting to gain access must adapt to the state of the network, which in turn changes the state. When characterizing the capacity of such systems, information and control theory very often come together. There is a vast and growing body of literature in different models for control and coordination in cognitive radios and the resulting information theoretic capacity of the medium. This is particularly relevant over the next 1-2 decades as increasing spectrum scarcity and possibly radical reform by spectrum governing agencies (e.g. the FCC, see the July 2012 PCAST report) on how spectrum is allocated may cause wireless networks to much more closely resemble a massive feedback control system. This also provides connections with Economics, see Section 12.
- **Control over communication channels:** When controlling single or multi-agent systems is the goal, especially over a communication channel, information theory has a natural role to play. In such settings, information theoretic tools are useful in understanding the resulting performance of the overall system. Specifically, the tradeoff between accuracy, delay and error in control can be better understood using a joint control-information theoretic perspective. Information theory has been used, in particular, to show impossibility results in obtaining desired control objectives within desired error and delay objectives.
- **Humans in the loop and brain-machine interfaces:** There is a broad range of future directions at the interface of control and information theory. Of particular interest is in understanding decision making with humans-in-the-loop, where internal biases, sentiments and other externalities play important roles. These factors are best captured using a combination of tools from control and information theory, where the goal is to model and influence decision making in institutions using a better understanding of the complex interplay between human agents. In addition, brain-machine interfaces, where machines directly take neural signals and convert them to actionable information, combining elements of control and information theory with neuroscience.

# 5 Neuroscience

Neuroscience is largely an endeavor to reverse-engineer the brain, an organ dedicated to processing information and acting upon it. Brains can extract detailed, high-level meaning from raw data streams and produce extremely robust motor control, using a large number of networked components communicating over relatively slow and noisy channels. Thus, neuroscience would seem ripe for the application of IT, and indeed classical information theory has been applied to neural systems by mapping the channel to different system components:

from sensory transduction machinery, to dendritic processes of a single neuron, to large-scale connections between brain areas, and even to the whole brain, from sensory input to behavioral output. Estimates of entropy and mutual information are dependent on particular measurement methods and limited data, which has hampered its wide application so far. Nevertheless, by measuring the difference between information-theoretic optimal performance and behaviorally measured performance, IT has been used as a tool to understand the relative contributions of various nervous system components to the generation of certain behaviors.

Connections and Future Directions
- **Sensory systems.** Like any physically realized system, the brain is bound by resource constraints, and on both evolutionary and developmental time scales, it has the opportunity to maximize – or at least improve – its use of available resources. The principle of maximizing channel information capacity — and the resulting function of histogram equalization — was used to explain the tuning characteristics of sensory systems (Laughlin 1981), and brought to neuroscience the realization that an understanding of the characteristics of real-world, or naturalistic sensory input, is a prerequisite to probing sensory nervous system function. Information theoretic quantities have also been used as objective functions for finding parsimonious representations and models of sensory processing. Information maximization, as a normative theory, has been successful at explaining many high-level functional characteristics of sensory systems.
- **Beyond Sensory systems and towards cognition.** Extending the *efficient coding hypothesis* (Barlow 1961) to other areas of neuroscience beyond sensory systems has been less successful to date. One well-appreciated reason for this is the divergence of information *value* (i.e. fitness-improving) from raw capacity. A theory of utility-adjusted information that preserves the mathematical tractability of classical information theory could allow extended use of IT-based analytical tools to understand central processing: principles of decision making, cognition, as well as output generation, i.e. motor control.
- **Brain as a Controller**. A prevalent theory of nervous system evolution is that the first function of nervous systems was to effect movement and other brain functions subsequently evolved from movement control, and as such, understanding motor control may elucidate general principles of brain function. An even stronger viewpoint, perhaps the intellectual descendant of cybernetics (Wiener), holds that cognitive function can only be understood in the context of closed-loop control. Optimal control theory has shown some success in producing normative models of neural motor control (Todorov), and transitively, connections between IT and control theory can be expected to be applicable to the neuroscience of motor control.
- **Brain as a Data Network**. Experiments continue to substantiate the modularity of brain function and the distributed nature of cognitive processes. Network information theory may be applicable to characterize the information flow between neurons and between areas of the brain.
- **Predictive Information**. Another normative model of brain function is that it is a prediction machine, justified by the observation that fitness improvement to an organism can only happen in the future. Predictive information — past data which improves knowledge of future probability distributions — may be a useful derived IT quantity that can span the gap from classical IT to biological function. (Bialek)

- **Consciousness as Information**. Perhaps the most ambitious extensions of classical IT toward neuroscientific applications has been the theory of integrated information: that consciousness itself can be quantified as the reduction of uncertainty conferred by interactions between connected parts of a system versus their unconnected state. While much work must be done both to provide methods for estimating such a quantity, and success is by no means assured, this line of research can be appreciated in its attempt to formulate a rigorous mathematical definition of the notoriously elusive but essential concept of consciousness.

# 6 Signal Processing

Signal processing in the context of information theory is largely about methods and algorithms to achieve information theoretic ideals. After all, information is almost invariably represented as a signal of some sort, thus making any implementation of information theoretic techniques inseparable from signal processing. Furthermore, there are several areas fundamental to signal processing that are of increasing interest in the information theory community.

Connections and Future Directions:
- **Interference alignment, interference cancellation, and other multiuser capacity-achieving techniques**. In classical multiuser IT channel models (e.g. broadcast channel, interference channel, etc.), the optimal transmitter and receiver solutions, when known, usually involve sophisticated interference cancellation and/or pre-cancellation techniques. These techniques have proved difficult to achieve in practice largely due to signal processing limitations such as low-resolution digital-to-analog convertors, insufficient channel information (due to estimation errors and/or latency), and algorithmic complexity. Thus, future advances in signal processing techniques and capabilities have considerable scope for approach capacity regions.
- **Full-duplex communication.** One universally assumed constraint (outside of certain information theory models like the relay channel) is that a transceiving device cannot send and receive information at the same time in the same frequency band. This long-standing wisdom is being energetically challenged recently with the advent of self-interference suppressing transceivers that can achieve previously unimagined separations between the transmit and receive chains in full duplex mode. Nominally doubling the bidirectional rate, such an advance would have many other implications on protocol design and would further enable information theoretic guidance, as in the case of relay or multihop networks where full duplex is a fundamental advantage.
- **Human information acquisition**. The bottleneck for information consumption is often the human sensory processing system. Modeling and characterizing the maximum amount of information that a human can consume per unit time in a fundamental way is an open problem. One example is lossy compression of video, where different types of information loss have radically different effects on the perception of video quality. Is there a characterizable upper bound (e.g. some extreme HD format) beyond which it is simply not possible to improve human perception of the quality by increasing the information rate? Similar questions could be formulated for other types of information consumption.

## 7 Statistics and Machine Learning

Extracting useful information from large amounts of data is a classical problem, but one that has taken on urgent importance in recent years, as a number of recent technological innovations have allowed the collection and storage of data at a previously unimaginable scale. Information theoretic methods have had significant impact historically on this topic, including hypothesis testing, sample complexity bounds, MDL theory and estimation theory. The new problems emerging from the era of big data are transforming areas like statistics, machine learning and data mining as tools for information processing.

Connections and Future Directions:
- **Computational efficiency.** Classical statistical methods were not typically concerned with the design of efficient algorithms. As data sets grow, faster methods are becoming more relevant even if they produce sub-optimal or approximate results. For example, the problem of sparse regression and the related area of compressed sensing position computational efficiency as a central parameter of importance. Recovering unknown sparse vectors observed through linear measurements was a classically studied problem when exhaustive search algorithms were considered. However, it was the recent theoretical developments that analyzed what polynomial time algorithms can actually achieve that created a renewed and increased interest in these problems. Message passing algorithms and convex optimization are two often used techniques for obtaining computationally efficient algorithms with theoretical guarantees.
- **High-dimensional statistical theory**. Traditional statistics typically assume that the number of observed samples is significantly higher than the number of features (or parameters) being estimated. In this classic asymptotic theory, the sample size $n$ grows to infinity while the number of parameters (dimensions) $p$ remains fixed. Motivated by the possibilities of collecting more detailed data, the emerging area of high-dimensional statistics studies problems when the number of samples $n$ is comparable or even smaller than the dimensionality $p$. In this setup, extracting meaningful information from noise is very challenging because the number of samples is small compared to the desired number of parameters that must be estimated. For this reason, it is necessary to assume some low-dimensional structure so that recovery is possible. Examples of low-dimensional structure include recovery of sparse vectors, low-rank matrices or manifolds. Classical statistical methods are not applicable in this regime, and thus several new methods are emerging. Information theoretic techniques have already been influential in this emerging area, for example in obtaining bounds on sample complexity.

## 8 Genomics and Molecular Biology

Information is central to biology, most obviously because DNA stores genetic information. The bulk of the results in information theory, however, were developed in the context of engineered systems, whereas biological systems have are the result of evolution. Information, communications and networks are undeniably part of biological systems. Information theory has made a tremendous impact to bioinformatics by introducing to this discipline the edit distance and accompanying dynamic programming algorithms for computing it efficiently. The edit (Levenshtein) distance is at the core of all sequence alignment algorithms, used for assessing similarities of genomic and proteomic data. As such, it is used in almost every branch of comparative biology, for reconstructing phylogeny and identifying potential disease genes. Unfortunately, although our community has ventured into genomics and molecular biology, information theory still does not have a strong presence in life sciences. Some examples of information-theoretic tools that may be of significant use in bioinformatics include (i) graphical models, such as factor graphs; (ii) the Viterbi and BP algorithm (as well as other decoding methods) for state likelihood estimation in DNA sequencing systems, gene and protein regulatory networks; (iii) information-theoretic methods for analyzing fundamental performance limits of next generation DNA sequencing devices and mass spectrometry systems; (iv) information-bottleneck methods for classification; (v) Statistical tests such as expectation maximization and hypothesis testing. One of the key challenges in biological applications is formulating models that are both empirically justified and amenable to information theoretic tools. Therefore, we need to use broader information theoretic tools and notions and develop statistical approaches tailored to the systems and data from biology, not the ones that have been specifically designed for engineered systems.

Two main families of challenges in application of IT in biology are (1) statistical averaging issues and (2) modeling issues. The former includes revisiting several central assumptions in most information theoretic analyses. For example, because the heterogeneity of phenotypes is a rule in biology, rather than an exception (e.g. cell size and content, cell receptor types and numbers, organ size and blood-flow), averages may not be very important in explaining or predicting behaviors. Similarly, stationarity generally does not hold and sample sizes may be too small to invoke the law of large numbers, and so time-averaged results must be interpreted cautiously. As far as modeling issues, key issues include (i) behavior of noise (often resulting in state changes, deletions/insertions, reorderings, rather than being additive and/or Gaussian) and (ii) robustness of models to imperfections. For example, many key results in classical information theory are based on idealizations (infinite block lengths, additive white Gaussian noise, etc.) that nevertheless hold quite well under many relaxations, but there is reason to believe the same will not be true in biology. For example, the Poisson channel, proposed to explain neural coding, is a fragile, sensitive, non-robust model. Any traces of memory in the system will obviate the results.

Connections with IT, and Future Directions:
- **Limits of DNA Processing (in presence of noise).** A key challenge here vs. traditional coding theory is that rather than "substitutions" (symbol errors or erasures caused by noise), a major source of errors in genomics, stems instead from deletions of desired symbols and insertions of unwanted symbols. Some sources of noise (e.g., misincorporation by polymerases) can be dependent on the DNA sequence context, and are therefore heterogeneous over sequences of interest in ways that aren't usually well characterized. This has been explored in the IT literature but is still a challenging and open topic.
- **Adapting IT Tools to Genomics.** Information-theoretic tools and metrics should be adapted to the specific aspects of genomic sequences, which are nonstationary and highly contextual. For example, the mutual information between two data series may prove particularly useful in genomics, compared to the correlation or other typical metrics used to compare two sequences. But it is hard to compute mutual information without knowing the joint distribution; tools for computing it empirically would likely prove very useful.
- **Evolutionary Biology.** Evolution is closely related to genomics, as evolution is largely about adaptation to an environment which is then stored in genes change over time. Information theory can be applied to many aspects of evolution including group behavior/cooperation, drug resistance and drug design, cognitive evolution, and as a framework for quantitatively describing evolution [Adami 2012].
- **Data Compression in Molecular Biology:** One of the important challenges in the 1000 Genome project was data compression and data transfer protocols. State-of-the art compression methods offer up to 10-20-fold reductions in high-throughput sequencing data, while the goal is 100-200-fold compression ratios. Further, recent IT work on searching in compressed data or computation with compressed data may have particularly promising applications in this area.
- **Bio-inspired algorithms for IT.** Many biological processes can be seen as algorithms that nature has designed to solve computational problems. Examples are evolutionary algorithms for large-scale problems in optimization, robust network design, and distributed computation. The decoding of LDPCs and turbo codes also have a biological flavor and complexity, e.g. the decoder operates as a distributed network with information transfer that obeys certain rules, but is difficult to quantify or predict.

# 9  Theoretical Computer Science

One traditional boundary line between information theory and theoretical computer science lies in whether computational complexity is a concern or not. Classical information theory is not concerned with efficiency since notions such as "capacity" or "source rate" are implementation agnostic. In contrast, theoretical computer science puts the computational/implementation burden at the center of its research. As we have noted above, many challenging open problems in information theory involve large interconnected systems, where even computing the capacity or other fundamental properties (let alone, achieving them) is very challenging.

Connections and Future Directions:
- **Communication Complexity and Information Theory.** Communication complexity is a sub-branch of theoretical computer science that studies the problem of determining the minimum amount of communication needed to perform a distributed computation. For example, two players A and B might each have an n-bit string and desire to compute a function of both strings. The question is to determine the minimum amount of communication required to perform to compute this function. Despite the natural similarity to problems studied in information theory, there has been limited cross-proliferation of ideas between these two areas. Several fundamental problems in communication complexity remain open and it is possible that information theoretic techniques might be applicable. The increasing relevance of distributed computation can possibly raise more interest in this area.

- **Locally Repairable Codes and Complexity Theory.** Coding for distributed storage is receiving increasing interest due to its applicability in large-scale data centers. The first families of distributed storage codes use network coding techniques to minimize the amount of communication during single storage node failures. More recently, distributed storage codes that minimize the number of nodes required to repair a single failure were introduced. These Locally Repairable Codes (LRCs) were recently deployed in production data centers but their theoretical properties have not been not fully characterized. A closely related concept in theoretical computer science is that of Locally Decodable Codes (LDCs). LDCs are used in obtaining hardness of approximation results and private information retrieval but not actual communication or storage. Their connections with LRCs and the applicability of algebraic coding theoretic techniques developed for LDCs in the design of LRCs are interesting future directions.
- **Index Coding and Graph Theory.** Index coding is a combinatorial problem that is becoming increasingly important in information theory. Recent results established that all wired network coding problems can be reduced to index coding, hence establishing the tremendous generality of the model. Further, graph theorists have shown the difficulty of index coding and the connections to other graph invariants like the chromatic and independence numbers. Historically, another related graph quantity sandwiched between independence and chromatic numbers, namely Shannon's zero-error graph capacity, motivated Lovasz to define his famous Theta function, possibly the first semidefinite relaxation studied for combinatorial optimization. SDP relaxations are now central tools in approximation algorithms used in theoretical computer science. It seems that index coding is a deep problem in the intersection between information theory, graph theory and theoretical computer science.

# 10 Physics

Information Theory strives to design and exploit correlations among random variables, with the hope to achieve specific goals: communicate a message through an unreliable channel, estimate a signal in noise, hide private information, and so on. Information Theory also focuses on very high dimensional probability (i.e. the message length or the number of degrees of freedom), where qualitatively new phenomena such as the possibility to communicate reliably over unreliable media at rates approaching the capacity. Meanwhile, for over a century physicists have developed conceptual tools to study very high-dimensional probability distributions with a different goal: explaining the macroscopic properties of matter stemming from its atomic components. In this process, they developed explanations of surprising behaviors, such as threshold phenomena like phase transitions.

Formal relationships between Physics and Information Theory have been known since the origins of IT. However, it was not until recently that tools from Physics have become impactful for Information Theory problems, which now attract significant attention from physicists. This cross-pollination will become increasingly fruitful.

Connections and Future Directions:

- **Mathematical tools** from theoretical and mathematical physics allow information theorists to characterize systems and study problems that cannot be tackled by classical methods. Examples range from the replica and cavity methods from spin glass theory that have been used to analyze modern error correcting codes and CDMA systems, to percolation theory that has played an important role in the analysis of connectivity in large wireless networks.
- **Intersection of theory and empirical/experimental research.** While some of these tools are fully rigorous, others have the status of mathematically sophisticated, albeit non-rigorous heuristics. If carefully used, and accompanied by intuition and independent checks, these heuristics can be used to sharpen the engineer's intuition. This suggests an altogether new style of research that stands halfway between the purely empirical and the completely rigorous. An example includes renormalization group theory. At the same time, information theorists can play an important role in further extending the utility and rigor of these tools.
**Physics-inspired algorithms and methods.** The same heuristics have provided important inspiration for new algorithms to tackle hard problems, for example those involving networks or iterative processes. Examples are abundant, in particular in connection with graphical models, belief propagation and sparse graph codes. As above, ideas from physics are particularly interesting in that they often offer an "orthogonal" line of attack on difficult problems that have been extensively studied by classical methods.

## 11 Economics and Finance

In spite of the prominent role that information plays in the economy, few intellectual connections have been made between information theory (IT) and economic theory. This is an area of interaction that is ripe for further exploration, particularly within the microeconomic context.

Connections and Future Directions:
- **Investment Theory.** The connections between IT and investment theory [Kelly, Latane, Cover, and others] are well established. Long-term expected capital growth can be maximized with the log-optimal strategy, i.e. maximizing the expected log of the return at each period. This is a generalization of the Kelly gambling rule. The log-optimal rule is optimal in expected ratio at every period and yields asset prices that correspond to data better than the classical CAPM model. Furthermore, the log-optimal rule can be used to price derivative assets, leading directly to the celebrated Black-Scholes equation.
- **Rationality.** Since the 1950s, the bounded rationality school of thought (work by Simon, Marshak, Radner, and others) has challenged the standard rational decision-making paradigm in economics. Rationality can be constrained by the cost of observation, computation, memory, and communication. Rationality can further be constrained by more inherent factors such as inconsistency, ambiguity, vagueness, unawareness, and failure of logical omniscience. This raises fundamental questions concerning the role of information and computational limits in economic theory.
- **Rational inattention** (RI) models (by Sims and others) focus on the idea that people's abilities to translate external data into action are constrained by an information processing capacity, measured by mutual information. RI models update classical rational expectations models within macroeconomics and finance, and imply pervasive inertial and erratic economic behavior.
- **Economic Value of Information**. Where information influences economic action, one can assign an objective value to the information, namely the difference in economic reward of an informed action over an uninformed one. In the context of decision making where the value of information is measured in monetary terms rather than bits, one can derive various relationships among information value quantities, which are generalizations of basic information inequalities (work by Marshak, Howard, DeGroot, and others).
- **Optimal Pricing with Limited Information.** Recently, a connection has been established between quantization in data compression and the problem of optimal pricing with limited information in microeconomics (work by Bergemann, Yeh, and others). Scalar and vector quantization techniques are used to solve open problems concerning the optimal finite pricing menus for welfare and revenue maximization, respectively, within a monopolistic setting. In particular, the use of vector quantization leads to important conclusions concerning optimal product bundling.
- **Information economics** (work by Arrow, Stiglitz, Akerlof, Spence, and others) is a major area within economics which has addressed important shortcomings of neoclassical economic theory. Information economics (IE) shows that information asymmetries in the economy can lead to phenomena which cannot be explained by neoclassical theory (which assumes perfect information). These include adverse selection, moral hazard, the absence of markets, asset price volatility, and persistent unemployment. Information theory has thus far not played a role in IE. This is a potential future direction.